\documentclass[twocolumn,preprintnumbers,amsmath,amssymb,superscriptaddress]{revtex4-1}

\usepackage{graphicx}
\usepackage{bm}
\usepackage{color}
\usepackage{ulem}
\usepackage{multirow}

\newcommand{\dd}{{\rm d}} 

\begin{document}

\title{Dissipation-driven emergence of a soliton condensate in a nonlinear electrical transmission line}
\author{Loic Fache}
\affiliation{Univ. Lille, CNRS, UMR 8523 - PhLAM -
  Physique des Lasers Atomes et Mol\'ecules, F-59 000 Lille, France}
\author{Hervé Damart}
\affiliation{Univ. Lille, CNRS, UMR 8523 - PhLAM -
  Physique des Lasers Atomes et Mol\'ecules, F-59 000 Lille, France}
\author{Fran\c{c}ois Copie}
\affiliation{Univ. Lille, CNRS, UMR 8523 - PhLAM -
  Physique des Lasers Atomes et Mol\'ecules, F-59 000 Lille, France}
\author{Thibault Bonnemain}
\affiliation{Department of Mathematics, King’s College, London, United Kingdom}
\author{Thibault Congy}
\affiliation{Department of Mathematics, Physics and Electrical Engineering, Northumbria University, Newcastle upon Tyne, NE1 8ST, United Kingdom}
\author{Giacomo Roberti}
\affiliation{Department of Mathematics, Physics and Electrical Engineering, Northumbria University, Newcastle upon Tyne, NE1 8ST, United Kingdom}
\author{Pierre Suret}
\affiliation{Univ. Lille, CNRS, UMR 8523 - PhLAM -
  Physique des Lasers Atomes et Mol\'ecules, F-59 000 Lille, France}
\author{Gennady El}
\affiliation{Department of Mathematics, Physics and Electrical Engineering, Northumbria University, Newcastle upon Tyne, NE1 8ST, United Kingdom}
\author{St\'ephane Randoux}
\email{stephane.randoux@univ-lille.fr}
\affiliation{Univ. Lille, CNRS, UMR 8523 - PhLAM -
  Physique des Lasers Atomes et Mol\'ecules, F-59 000 Lille, France}

\date{\today}

\begin{abstract}
We present an experimental study on the perturbed evolution of Korteweg-deVries soliton gases in a weakly dissipative nonlinear electrical transmission line. The system's dynamics reveal that an initially dense, fully randomized, soliton gas evolves into a coherent macroscopic state identified as a soliton condensate through nonlinear spectral analysis. The emergence of the soliton condensate is driven by the spatial rearrangement of the systems's eigenmodes and by the proliferation of new solitonic states due to nonadiabatic effects, a phenomenon not accounted for by the existing hydrodynamic theories.  
\end{abstract}


\maketitle

In the seminal article from 1965, Zabusky and Kruskal discovered stability and interaction properties of solitons in numerical simulations of the Korteweg-deVries (KdV) equation which they had obtained as a long-wave approximation of the Fermi–Pasta–Ulam-Tsingou oscillator chain \cite{Zabusky:65}. Two years later, Gardner {\it et al.} introduced the inverse scattering transform (IST) method, which provides an analytical framework to solve initial value problems for the KdV equation with rapidly decaying initial conditions \cite{Gardner:67}. IST theory was then generalized to other nonlinear dispersive equations of physical relevance, such as the one-dimensional nonlinear Schr\"odinger equation (1D-NLSE) \cite{Zakharov:72,Novikov_book,yang2010nonlinear}, which is a universal model for the description of many physical systems.

Solitons play a fundamental role in nonlinear physics due to their remarkable property of retaining shape, amplitude, and velocity upon interactions with other solitons \cite{Newell_book_solitons,Miura:76,Drazin_book,Copie:23}. This is reflected in IST theory by the fact that solitons represent normal modes of integrable nonlinear partial differential equations. The IST formalism can be seen as a canonical transformation which carries the physical coordinates, in which the integrable equations are originally given, to new (IST) coordinates which are action-angle variables. Within the IST formalism, the amplitude and velocity of each individual soliton are encoded into a discrete eigenvalue $\eta_k$, which represents an action variable that is time-invariant -- the so-called isospectrality property of integrable evolution \cite{Newell_book_solitons,Miura:76,Drazin_book}.

From the earliest days of IST theory, it was understood that the decomposition of the nonlinear field into a basis of normal modes represents a powerful tool applicable to the study of perturbed soliton evolutions relevant to physical applications where purely integrable dynamics are unlikely \cite{Karpman:77,Kaup:78,Kivshar:89}. The evolution of a KdV soliton under the action of weak dissipation is a prototypical example where the effects of a small perturbation on an integrable equation have been studied extensively \cite{Madsen:69,Grimshaw:70,Johnson:73,Kaup:78,Kodama:81,Kivshar:89,Grimshaw:94,Johnson:94,Ko:78,Knickerbocker:80,Knickerbocker:85,GEl:02}.

On the other hand, the question of emergent, large-scale behaviors in integrable or nearly-integrable many-body systems, such as soliton gases (SGs), is one of the main focuses of present-day experimental and theoretical physics \cite{Schemmer:19,Castro:16,Denardis:18,Ruggiero:20,Redor:19,Marcucci:19,Suret:20,Suret:23,Fache:24,Mossman:24,Delvecchio:20,Congy:24}. The out-of-equilibrium evolution of SGs can be investigated using the complementary tools of the nonlinear spectral kinetic theory of SGs \cite{Zakharov:71,GEl:05, GEl:20} and generalized hydrodynamics (GHD), the hydrodynamic theory for many-body quantum and classical integrable systems \cite{Doyon:18,doyon_lecture_2020,Castro:16,Bertini:16,Bonnemain:24}. The correspondence between these two hydrodynamic theories has been recently established in ref. \cite{Bonnemain:22} for KdV SGs. 

In this Letter, we describe experiments that explore the perturbed evolution of KdV SGs in a weakly dissipative nonlinear medium. We conduct our experiments in an electrical nonlinear transmission line (NLTL), where diffusion plays a perturbative role, and the dynamics are accurately captured by the KdV-Burgers (KdVB) equation. Our experiments reveal that an initially dense KdV SG, manifested as  a strongly oscillating random field, slowly evolves into a macroscopically coherent structure consisting of a broad smooth nonlinear rarefaction wave connected at  its leading edge to an oscillating diffusive-dispersive shock wave \cite{GEl:17}. From the perspective of the kinetic theory of SG, the formed composite nonlinear wave represents a {\it soliton condensate}, a critically dense SG  whose macroscopic properties are dominated by soliton interactions while the individual soliton dynamics are not discernible \cite{GEl:20, Congy:23}.  Soliton condensates have peculiar spectral properties which we identify by analyzing the IST eigenfunctions (or normal modes) composing the resulting wavefield. As was shown in \cite{Congy:23} the integrable KdV SG dynamics do not permit a spontaneous formation of a soliton condensate from a regular (non-condensate) SG evolution. Contrastingly, as we demonstrate, the formation of a soliton condensate from a non-condensate initial state in a {\it perturbed integrable system} becomes possible owing to a complex nonadiabatic process of the generation of numerous new solitonic modes, a phenomenon not accounted for by the existing hydrodynamic or kinetic theories of SG.

\begin{figure}[!]
  \includegraphics[width=8.5cm]{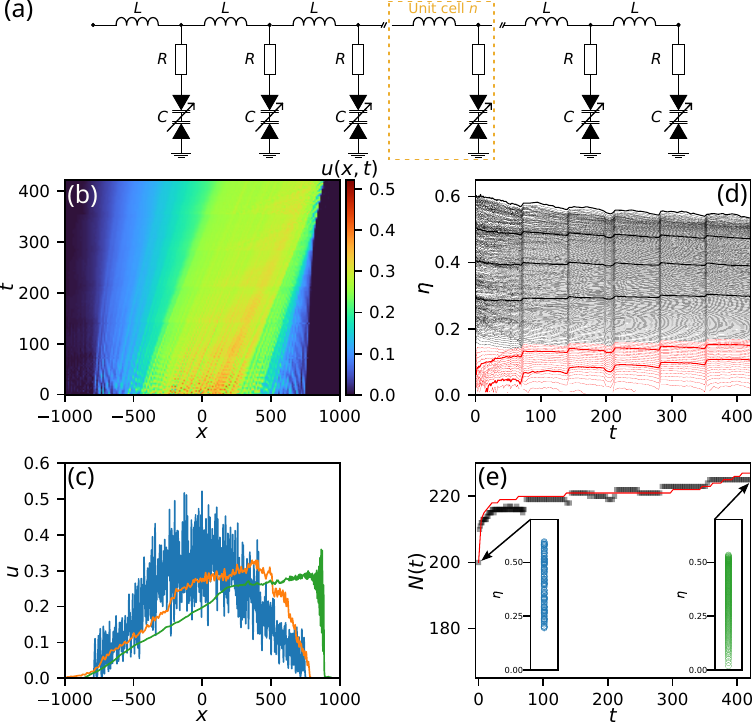}  
  \vspace{-5pt}
  \caption{Experiments. (a) Schematic representation of the NLTL. (b) Space-time plot showing the emergence of the soliton condensate in the NLTL. (c) Measured SG at $t=0$ (blue line), $t=175$ (orange line), $t=420$ (green line). (d) Time evolution of the discrete eigenvalues measured during the evolution of the SG. The red flow is associated to newly generated eigenvalues arising from the excitation of the continuous spectrum. (e) Time evolution of the number $N(t)$ of discrete eigenvalues measured during the evolution of the SG. The left (resp. right) inset shows the discrete IST spectrum at $t=0$ (resp. $t=420$). The red line in (e) is obtained from numerical simulations of the experiment with $\epsilon=0.1$. 
}
\end{figure}

Our experimental setup, shown schematically in Fig. 1(a), is an NLTL with an architecture similar to that used in previous investigations of KdV soliton propagation \cite{Remoissenet_book,ricketts_book_18,Jager:81,Elizondo:15}. It consists of a ladder network of $160$ identical lumped inductor-varactor sections with inductance $L$ and varactor capacitance $C(V)$. The varactors are commercial components (Varicap) for which $C(V)$ is a linearly-decreasing function of the applied voltage $V$. The varactors have a small series resistance $R$, which results into frequency-dependent NLTL losses. Hence, propagation in the NLTL in the long-wavelength limit is not described by the KdV equation but by the KdVB equation \cite{ricketts_book_18}:
\begin{equation}\label{eq:kdvb}
  u_t + 6 u u_x +u_{xxx} = \epsilon u_{xx}, 
\end{equation}
where $u(x,t)$ represents the normalized electrical voltage, see Supplemental material for details about the experimental setup and the normalization. The strength $\epsilon$ of the Burgers term in Eq. (\ref{eq:kdvb}) is determined by the resistance $R$ of the varactors. In our experiments, the value of $\epsilon$ is $0.1$. 

The initial condition $u_0(x)=u(x,t=0)$ used in our experiment is a dense SG numerically synthesized as a $N$-soliton solution of the KdV equation using the Darboux transformation method \cite{Congy:23,Liao:96,Perego:24}, see Supplemental material for details. The discrete IST spectrum of the initial condition is composed of $N=200$ discrete eigenvalues  $\eta_k$ that are randomly distributed between $0.2$ and $0.6$ in a uniform way, see Fig. 1(e). As shown in Fig. 1(c), the initial field measured at the first $LC$ cell of the NLTL is a random oscillating field in which individual solitons cannot be discerned due to their strong overlap and interactions. 

Measuring the voltage along the NLTL, we have built a space-time plot revealing the evolution of the SG up to $t=420$, see Fig. 1(b). In the absence of the diffusion term ($\epsilon=0$), the dense SG initially localized in space would have evolved into a rank-ordered expanding soliton train, see ref. \cite{Karpman:67} and Fig. 3(g). Contrastingly, we observe that the presence of the small diffusion term prevents the dense SG from reaching a diluted state where individual solitons separate, see Fig. 1(b)(c). Physically, weak dissipation promotes soliton interactions, leading to the emergence at long time ($t \sim 400$) of a broad, smooth wavefield with a coherent oscillatory structure identified as a diffusive-dispersive shock wave \cite{GEl:17,ricketts_book_18}.

Using the perspective of the IST method, we can gain insights into the features observed in the experiment. It is well known that the IST spectrum of the solutions of the KdV equation (Eq. (\ref{eq:kdvb}) with $\epsilon=0$) can be obtained by solving the linear Schr\"odinger equation for an auxiliary function $\psi(x,t)$
\begin{equation}\label{eq:schrodinger}
 \psi_{xx} + (u(x,t) -\lambda) \psi=0,
   \end{equation}
where the KdV field $u(x,t)$ plays the role of a potential with $t$ being a parameter \cite{Newell_book_solitons,Miura:76,Drazin_book}. For the fundamental soliton solution of the KdV equation  $u_k(x,t)=2 \eta_k^2 \, \text{sech}^2(\eta_k(x-4\eta_k^2t-x_{0k}))$, Eq. (\ref{eq:schrodinger}) admits only one eigenmode that is parameterized by one real discrete eigenvalue $\lambda_k=-\eta_k^2$ and by one normalized real eigenfunction $\psi_k(x,t)$ that is localized in space ($|\psi_k(x,t)| \rightarrow 0$ for $|x| \rightarrow \infty$ and $\int_{- \infty}^{+\infty} \psi_k^2(x,t) \, dx =1$) \cite{Drazin_book}. Importantly, any $N$-soliton solution of the KdV equation with purely discrete IST spectrum can be represented using the following reconstruction formula \cite{Gardner:74,Miura:76,Newell_book_solitons}
\begin{equation}\label{eq:inverse}
  u(x,t)= 4 \sum_{k=1}^N \eta_k \, \psi_k^2(x,t).
\end{equation}

Using a perturbative approach, where the Burgers term in Eq. (\ref{eq:kdvb}) is considered to be small, we computed the discrete eigenvalues $\eta_k$ associated with the experimentally measured field $u(x,t)$ by solving Eq. (\ref{eq:schrodinger}) at various times $t$. If the dynamics were governed by the KdV equation, one would expect an isospectral evolution, wherein the $N=200$ eigenvalues specified by our initial condition would remain unchanged over time. Instead, we observe that the eigenvalues distribute along two distinct ``flows'', see Fig. 1(d). In the first flow (black lines in Fig. 1(d)), the eigenvalues associated with the eigenmodes composing the initial SG slowly decay with time, as expected from the soliton perturbation theory \cite{Karpman:77,Kaup:78,Kivshar:89}. The second flow (red lines in Fig. 1(d)) consists of the newly formed discrete eigenvalues progressively emerging during the evolution. These new eigenvalues correspond to the generation of small-amplitude solitons whose number $N$ grows with time, see Fig. 1(e) showing that $N$ grows from $N(t=0)=200$ to $N(t=420)=225$. 

\begin{figure}[!]
  \includegraphics[width=8.5cm]{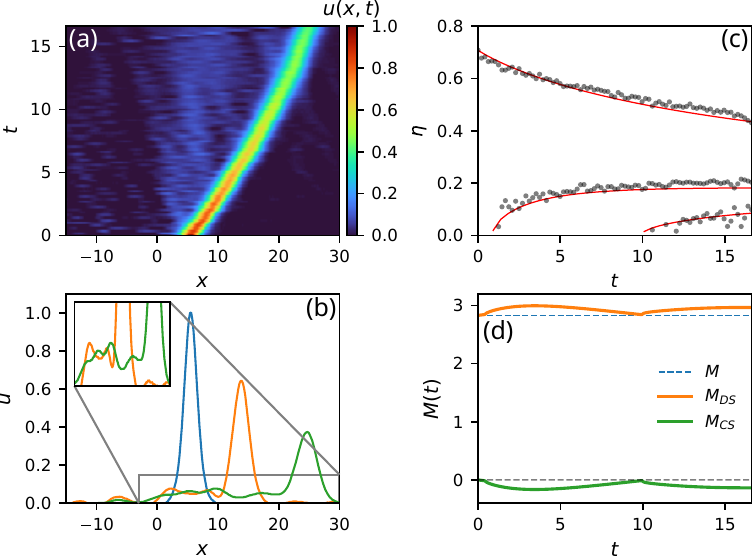}  
  \vspace{-5pt}
  \caption{Experiments. (a) Space-time evolution of a solitary wave in the NLTL. (b) Measured solitary wave at $t=0$ (blue line), $t=5.8$ (orange line), $t=16.6$ (green line). (c) Time evolution of the discrete eigenvalues measured during the evolution of the solitary wave. The red lines in (c) are obtained from numerical simulations of the experiment ($\epsilon=0.1$). (g) Numerical simulations showing the oscillatory evolution of the masses $M_{DS}$ and $M_{CS}$ of the discrete and continous spectra. 
}
\end{figure}

The described complex process of the proliferation of new soliton states represents a counterpart of a simpler phenomenon observed in the individual evolution of a solitary wave in the NLTL. As seen in Fig. 2(a)(b), the amplitude and velocity of the soliton, taken as the initial condition, slowly decay during the propagation in the NLTL. This gradual evolution of the solitary wave is accompanied by the formation of a trailing shelf, which is analogous to a prominent feature of the propagation of solitary waves in shallow water with slowly varying depth \cite{Knickerbocker:85, Kaup:78, Newell_book_solitons}.  Nonlinear spectral analysis, performed by solving Eq. (\ref{eq:schrodinger}) for the measured field $u(x,t)$, reveals that the discrete eigenvalue linked to the initial condition gradually decays over time, while two new discrete eigenvalues are created during the shelf formation, as seen in Fig. 2(c).

This is a non-adiabatic process because, although the soliton's amplitude changes slowly, the shelf's extent does not, resulting in an $O(1)$ mass of the shelf despite its small amplitude. Consequently, the solitary wave excites radiative modes (associated with the continuous IST spectrum), potentially leading to the creation of new solitonic eigenmodes \cite{Kaup:78, Newell_book_solitons,Wright:80, GEl:02}. In our experiment described by the KdVB equation, the excitation of the continuous spectrum by the solitary wave is observed through opposite oscillations between the masses $M_{DS}$ and $M_{CS}$ of the discrete and continuous spectra, such that the total mass $M = \int u(x,t)  \dd x = M_{DS} + M_{CS}$ is conserved \cite{Wright:80}, see Fig. 2(d) and Supplemental Material for details. 

\begin{figure*}[!]
  \includegraphics[width=\textwidth]{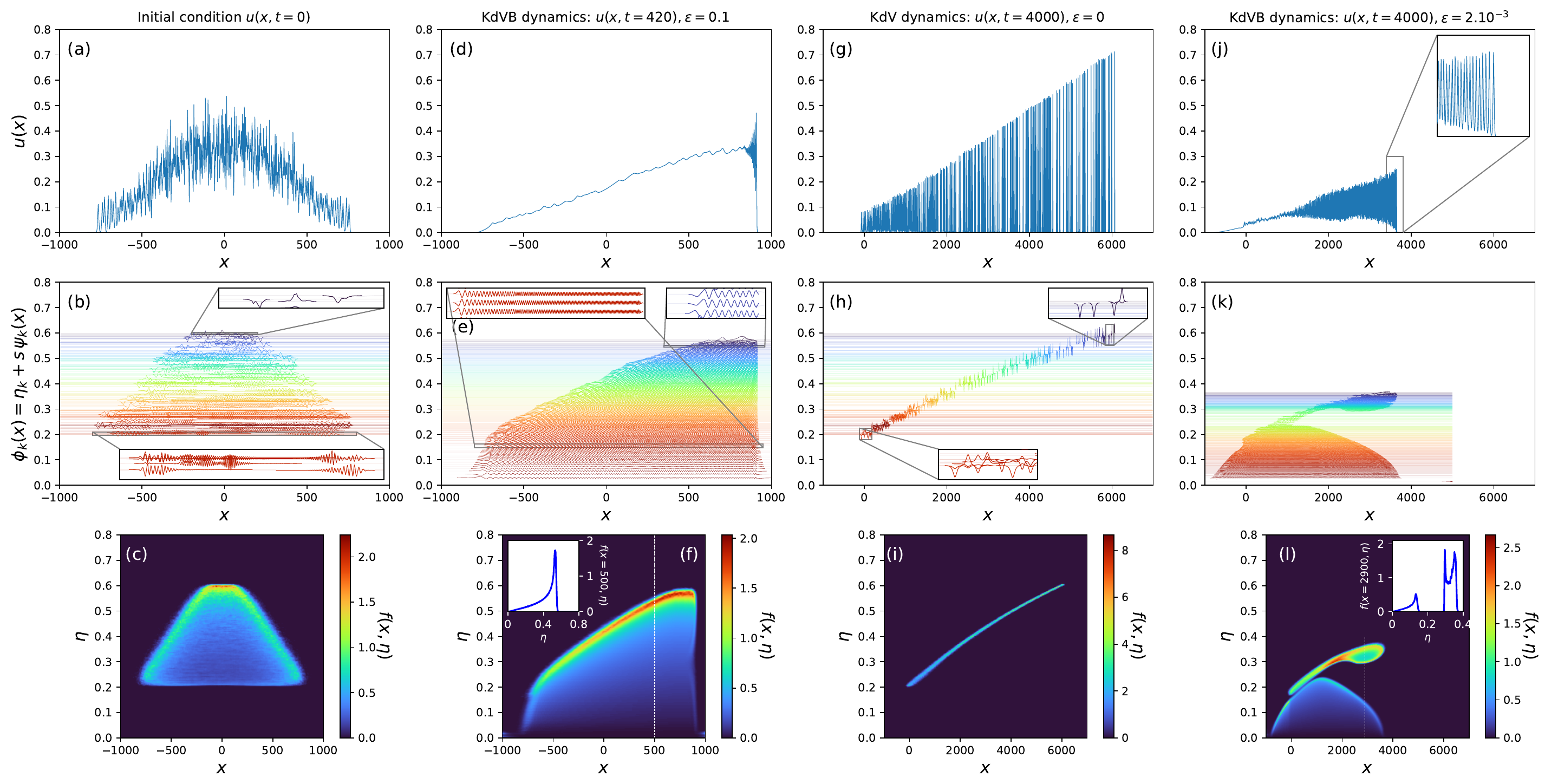}    
  \vspace{-20pt}
  \caption{Numerical simulations. (a) SG used as initial condition in the experiment. (b) Spatial distribution of the discrete eigenmodes represented as $\phi_k(x)=\eta_k+ s \, \psi_k(x)$, where $\eta_k$ (resp. $\psi_k(x)$) ($k=1,..200$) represent the eigenvalues (resp. eigenfunctions) computed by solving Eq. (\ref{eq:schrodinger}). (c) DOS of the SG shown in (a). (d), (e), (f) Same as (a), (b), (c) but for a simulation of Eq. (\ref{eq:kdvb}) at $t=420$ with $\epsilon=0.1$. (g), (h), (i) Same as (a), (b), (c) but for a simulation of Eq. (\ref{eq:kdvb}) at $t=4000$ with $\epsilon=0$. (j), (k), (k) Same as (a), (b), (c) but for a simulation of Eq. (\ref{eq:kdvb}) at $t=4000$ with $\epsilon=2.10^{-3}$.  The inset in (f) (resp. in (l)) shows the DOS measured at $x=500$ (resp. $x=2900$).
}
\end{figure*}

Using nonlinear spectral analysis, we now examine how these nonadiabatic effects impact the evolution of the perturbed KdV SG. Fig. 3(a) shows the SG used as initial condition in our experiment. The associated distribution of the discrete eigenvalues $\eta_k$ and eigenfunctions $\psi_k(x)$ parameterizing this SG is plotted in a single representation showing $\phi_k(x)=\eta_k+ s \, \psi_k(x)$ ($k=1,..200$), see Fig. 3(b). The eigenvalues and eigenfunctions are computed by solving numerically Eq. (\ref{eq:schrodinger}) for $u(x,t)=u_0(x)$. We  introduced a scaling factor, $0<s \ll 1$, to prevent overlap among the eigenfunctions in Fig. 3(b). 

As shown in the insets of Fig. 3(b), the degree of localization in space of the eigenfunctions composing the random initial condition $u_0(x)$ depends of their associated eigenvalues. Nonlinear discrete eigenmodes with the largest eigenvalues ($\eta_k \sim 0.6$) are more localized in space than the modes with the smallest eigenvalues ($\eta_k \sim 0.2$). Taking $u_0(x)$ as the initial condition and solving numerically Eq. (\ref{eq:kdvb}) for $\epsilon=0.1$, we obtain the field shown in Fig. 3(d), in good quantitative agreement with the experimental measurements. Fig. 3(e) reveals that the distributions of the eigenvalues and of the eigenfunctions composing the field have been profoundly modified by the nonlinear evolution. In addition to the creation of new eigenfunctions with spectral parameters between $0$ and $0.2$, all the eigenfunctions composing the wavefield at $t=420$ have lost their random structure and adopted similar uniform oscillatory shape. 

This large-scale behavior is in sharp contrast with that observed in the absence of diffusion, for $\epsilon=0$. As shown in Fig. 3(g), for a pure KdV evolution, a rank-ordered soliton train forms at long evolution time, as a result of an isospectral evolution where the eigenfunctions composing the nonlinear wave field become spatially localized as solitonic peaks, see Fig. 3(h). An intermediate situation between the experimental case ($\epsilon=0.1$ in Fig. 3(d)) and the purely integrable case ($\epsilon=0$ in Fig. 3(g)) is observed for a very small diffusion parameter ($\epsilon=2.10^{-3}$). In this situation, the SG exhibits a slowed-down expansion, in which a random soliton train propagates over a non-zero background, see Fig. 3(j). Fig. 3(k) shows that the associated eigenfunctions are divided into two classes: the localized modes and the delocalized modes, the latter originating from the excitation of the continuous spectrum.

Now we show that the squared eigenfunctions $\psi_k^2(x,t)$ are directly associated with the density of states (DOS) $f(\eta_k;x,t)$ in the SG spectral kinetic theory, revealing their fundamental role in this framework. The SG spectral kinetic theory is a statistical theory describing the transport in space and time of the DOS. The DOS $f(\eta;x,t)$ is defined as the joint distribution over the spectral eigenvalues and the soliton positions, so that $f(\eta;x,t) \dd \eta \dd x$ represents the number of soliton states found at time $t$ in the element of the phase space $[\eta,\eta+\dd \eta] \times [x,x+\dd x]$ \cite{GEl:20,GEl:21,Suret:20,Suret:23,Fache:24}. In particular, knowledge of the DOS provides the statistical moments of the field $u(x,t)$, including its mean value \cite{GEl:21,Bonnemain:22}:
\begin{equation}\label{eq:mean}
  \langle u \rangle (x,t)= 4 \int_0^1 \eta f(\eta;x,t) \dd \eta. 
\end{equation}
The brackets in Eq. (\ref{eq:mean}) denote ensemble averaging over a large number of realizations of the SG. 

By comparing Eq. (\ref{eq:inverse}) with Eq. (\ref{eq:mean}), one can infer that the DOS can be computed using the equality
\begin{equation}\label{eq:dos}
f(\eta_k;x,t) \dd \eta = \sum_{j=k}^{k+\Delta k} \langle \psi_j^2(x,t) \rangle, 
\end{equation}
where $k$ and $k+\Delta k$ are integer indices bounding the discrete summation over the squared eigenfunctions associated with eigenvalues between $\eta_k$ and $\eta_k+ \dd \eta$. Equation (\ref{eq:dos}) provides a {\it local} expression for the DOS of a SG, akin to formulas used in other branches of physics such as electron microscopy \cite{Abdajo:08,Tersoff:85}. 

Fig. 3(c), (f), (i), (l) show the DOS of the SGs from our numerical simulations. These plots were created using Eq. (\ref{eq:dos}) and $1000$ realizations of each SG shown in Fig. 3(a), (d), (g), (j). Fig. 3(c) shows that the DOS of the initial SG is not uniform in space. The fact that the KdV evolution leads to the emergence of a rank-ordered soliton train (Fig. 3(g)), with eigenfunctions becoming localized in space (Fig. 3(h)), translates into the DOS being localized around the curve shown in Fig. 3(i).

Conversely, the SG formed in the experiment under the perturbative influence of diffusion exhibits a very specific DOS, locally described by the Weyl distribution $f(x,\eta)=\eta/(\pi\sqrt{\lambda_1^2(x)-\eta^2})$ \cite{Congy:23}, with the function $\lambda_1(x)$ being slowly modulated in space, see Fig.~3(f). This DOS corresponds to the peculiar type of SG known as a genus zero soliton condensate in the spectral kinetic theory of SG \cite{Congy:23, Kuijlaars:21, Gelash:21, GEl:20}. Here this soliton condensate can be seen as the SG of minimal entropy or of zero variance \cite{Bonnemain:22}.

For a diffusion strength smaller than the experimental one ($\epsilon = 2.10^{-3}$), the DOS at a given position $x > 0$ is supported on two disjoint spectral intervals, see Fig.~3(l). The corresponding coherent oscillatory structure, identified above as a dispersive-diffusive shock wave, is also characterized by the soliton condensate DOS within the SG spectral framework, however, now it is the genus one condensate, see Supplemental Material and \cite{Congy:23} for details. As shown in \cite{Congy:23}, the transitions from a non-condensate to the condensate state and vice versa are not possible in an integrable system. Our experiment demonstrates that soliton condensate can spontaneously emerge in a nearly integrable physical system due to the presence of a perturbative dissipative effect. This observation suggests the possibility of performing further experiments equivalent of Joule expansion, by evolving a SG according to KdVB equation and suddenly setting the dissipation parameter $\epsilon$ to zero. 

In summary, we have reported the emergence of a soliton condensate in a NLTL, a nearly integrable physical system described by the KdVB equation. This phenomenon arises from the dissipation-driven rearrangement of solitonic eigenmodes within the system, accompanied by the excitation of the continuous spectrum, resulting in the formation of new soliton states. The coupling between solitons and continuous spectrum radiation has been recently highlighted in deep-water models less restrictive than the integrable 1D-NLSE \cite{Gelash:24}. Notably, current hydrodynamic theories like GHD have yet to incorporate the excitation of the continuous spectrum by a statistical ensemble of solitons. We anticipate that our work will inspire further research in this area.

\begin{acknowledgments}
This work has been partially supported  by the Agence Nationale de la Recherche  through the StormWave (ANR-21-CE30-0009) and SOGOOD (ANR-21-CE30-0061) projects, the LABEX CEMPI project (ANR-11-LABX-0007), the Ministry of Higher Education and Research, Hauts de France council and European Regional Development Fund (ERDF) through the Nord-Pas de Calais Regional Research Council and the European Regional Development Fund (ERDF) through the Contrat de Projets Etat-R\'egion (CPER Photonics for Society P4S). The authors would like to thank the Centre d'Etudes et de Recherche Lasers et Application (CERLA) for technical support and the Isaac Newton Institute for Mathematical Sciences for support and hospitality during the programme ``Dispersive hydrodynamics: mathematics, simulation and experiments, with applications in nonlinear waves''. TB was supported by the Engineering and Physical Sciences Research Council (EPSRC) under grant EP/W010194/1. GE's and GR's work was also supported by the EPSRC, grant number EP/W032759/1. GR thanks the Simons Foundation for partial support. 

\end{acknowledgments}





\bibliographystyle{apsrev4-1}

\pagebreak
\newpage

\renewcommand{\theequation}{S\arabic{equation}}
\setcounter{equation}{0}
\onecolumngrid


\renewcommand{\theequation}{S\arabic{equation}}
\renewcommand{\thefigure}{S\arabic{figure}}

\vspace{14cm}

\begin{center}
{\bf Supplemental material for : \\"Dissipation-driven  emergence of a soliton condensate in a nonlinear electrical transmission line"}\\
\end{center}

\begin{center}
    Loic Fache,$^1$ Hervé Damart,$^1$ Fran\c{c}ois Copie,$^1$ Thibault Bonnemain, $^2$, Giacomo Roberti,$^3$, Thibault Congy,$^3$ Pierre Suret,$^1$ Gennady El,$^3$ St\'ephane Randoux$^1$
\end{center}

\begin{center}
  {\it $^1$ Univ. Lille, CNRS, UMR 8523 - PhLAM -Physique des Lasers Atomes et Mol\'ecules, F-59 000 Lille, France

    $^2$ Department of Mathematics, King’s College, London, United Kingdom
    
  $^3$ Department of Mathematics, Physics and Electrical Engineering, Northumbria University, Newcastle upon Tyne, NE1 8ST, United Kingdom
  }
\end{center}

The purpose of this Supplemental Material is to provide some details about the experimental setup and about the experimental methododology.  All equations, figures, reference numbers within this document are prepended with ``S'' to distinguish them from corresponding numbers in the Letter.\\

\tableofcontents

\section{\label{sec1} Description of the experimental setup and of its associated dynamical equations}

\begin{figure}[h]\label{figS1}
    \centering
    \includegraphics[width=\textwidth]{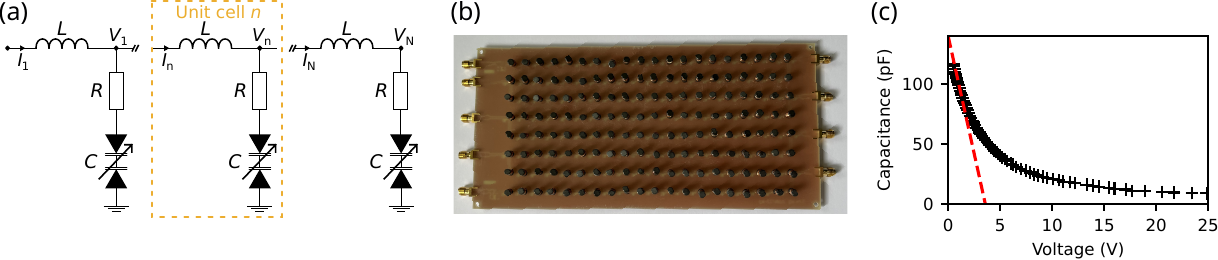}
    \caption{(a) Schematic representation of the experimental setup consisting into a ladder network of $160$ lumped inductor-varactor sections. The inductors have a constant inductance $L \simeq 15 \mu$H while the varactors have a capacitance that decreases with the applied voltage. The varactors have also a small but non-negligible series resistance $R \simeq 5 \, \Omega$. (b) Photograph of the experimental setup. (c) Typical characteristic of the varicap used in the NLTL showing the evolution of the capacitance $C$ as a function of the applied voltage. The characteristic is nearly linear for positive voltages below $\sim 3$ V. For this range of voltage, it can be approximated by $C(V_n)=C_0(1-2bV_n)$ with $C_0=140$ pF and $b=0.154$ V$^{-1}$.
      }
\end{figure}

Our experimental setup is the electrical nonlinear transmission line (NLTL) schematically shown in Fig. S1(a). It consists into a ladder network of $160$ lumped inductor-varactor sections with each individual inductance (KEMET SBCP-47HY150B) being $L \simeq 15 \mu$ H and each varactor capacitance $C(V_n)$ being a decreasing function of the voltage $V_n$ applied to it. The varactors are commercial components (Varicap: NXP-BB201). Fig. S1(c) shows the measured capacitance $C(V_n)$ as the function of the voltage $V_n$. For positive voltages below $\sim 3$ V, the capacitance decreases approximately linearly with the voltage and the curve plotted in Fig. S1(c) can be well approximated by $C(V_n)=C_0(1-2bV_n)$ with $C_0=140$ pF and $b=0.154$ V$^{-1}$. Finally and importantly, the varactors have a small but non-negligible resistance $R \sim 5 \, \Omega$. \\

The theoretical derivation of the dynamical equations describing the NLTL shown in Fig. S1(a) has been reported in details in ref. \cite{ricketts_book_18}. For the sake of clarity and completeness, we report here the main steps showing that nonlinear propagation in the NLTL in the long-wavelength limit is described by the KdV-Burgers equation. The nodal equations for the voltages and currents shown in Fig. S1(a) read: 
\begin{equation}\label{eq:kirch1}
    I_n - I_{n-1} = \frac{\partial Q(V_n) }{ \partial t }, 
\end{equation}
\begin{equation}\label{eq:kirch2}
    L \frac{\partial I_n}{\partial t} = \left[ V_{n-1} + R \frac{\partial Q(V_{n-1})}{\partial t} \right] - \left[ V_{n} + R \frac{\partial Q(V_{n})}{\partial t} \right],
\end{equation}
\begin{equation}\label{eq:kirch3}
    L \frac{\partial I_{n+1}}{\partial t} = \left[ V_{n} + R \frac{\partial Q(V_{n})}{\partial t} \right] - \left[ V_{n+1} + R \frac{\partial Q(V_{n+1})}{\partial t} \right],
\end{equation}

Combining Eqs. (\ref{eq:kirch1}), (\ref{eq:kirch2}), (\ref{eq:kirch3}), we obtain: 
\begin{equation}\label{eq:kirch4}
    L \frac{\partial^2 Q(V_{n})}{\partial t^2} = V_{n-1} + V_{n+1} - 2V_n + R \frac{\partial}{\partial t} \left[ Q(V_{n+1}) + Q(V_{n-1})- 2Q(V_n) \right]
\end{equation}

Changing $V_n(t)$ into $V(n,t)$ and Taylor-expanding the voltages $V(n,t)$ as
\begin{equation}\label{eq:taylor}
  V_{n \pm 1} = V \pm \frac{\partial V}{\partial n} + \frac{1}{2!}\frac{\partial^2 V}{\partial n^2}  \pm \frac{1}{3!}\frac{\partial^3 V}{\partial n^3} + \frac{1}{4!}\frac{\partial^4 V}{\partial n^4} \pm O(n^5),
\end{equation}
we obtain 
\begin{equation}\label{eq:kirch5}
  V_{n+1} + V_{n-1} - 2V_n = \frac{\partial^2 V(n,t)} {\partial n^2} + \frac{1}{12} \frac{\partial^4 V(n,t)} {\partial n^4} + O(n^5)
\end{equation}
Substituting the charge-voltage relation of the varactor $Q(V_n)=C_0(V_n-b V_n^2)$ in Eq. (\ref{eq:kirch4}), we find that propagation in the NLTL in the long-wavelength limit is described at leading order by the following partial differential equation \cite{ricketts_book_18}:
\begin{equation}\label{eq:pde}
   V_{tt} - b (V^2)_{tt}=v_0^2 \left( V_{nn} + \frac{V_{nnnn}}{12} \right) + \frac{R}{L} V_{nnt},
\end{equation}
where $v_0=1/\sqrt{LC_0}$. The subscripts in Eq. (\ref{eq:pde}) denote derivation with respect to $n$ and $t$. Following ref. \cite{ricketts_book_18} a small dimensionless parameter $\alpha$ is introduced for scaling linear and nonlinear effects. Changing $V$ into $\alpha V$, $R/L$ into $\alpha^{1/2} R/L$ and introducing the following new variables:
$$
  \begin{cases}
    s=\alpha^{1/2}(n+(\alpha-1)v_0 t),\\
    \tau=\alpha^{3/2} t,
  \end{cases}
$$

we obtain the following equation that describes the nonlinear propagation of waves in the NLTL in physical variables  
\begin{equation}\label{KdV_space}
\frac{\partial V}{\partial \tau} + v_0 \frac{\partial V}{\partial s} + b v_0 V \frac{\partial V}{\partial s} + \frac{v_0}{24}\frac{\partial ^3 V}{\partial s^3} = \frac{R}{2L} \frac{\partial^2 V}{\partial s^2}. 
\end{equation}
Considering that $V_{s} \sim -\frac{1}{v_0}V_{\tau}$, the equation describing the space-time evolution of the voltage along the LC oscillator chain reads:
\begin{equation}
    V_s + \frac{1}{v_0}V_{\tau} - \frac{b}{c_0}VV_{\tau} - \frac{1}{24 v_0^3}V_{\tau \tau \tau} - \frac{R}{2Lc_0^3}V_{\tau \tau} = 0.    
\end{equation}
Finally introducing the following normalized space and time variables together with the normalized field $u(x,t)$ 
\begin{equation}\label{normalization}
  \begin{cases}
    t=\frac{1}{3}\sqrt{b^3 V_0^3}s, \\
    x=-2 v_0 \sqrt{bV_0}\left(\tau -\frac{s}{v_0} \right), \\
    u = \frac{V}{V_0}, 
  \end{cases}
\end{equation}
we obtain the KdV-Burgers equation: 
\begin{equation}
    u_t+6uu_x+u_{xxx}=\epsilon u_{xx}, 
    \label{Kdv_usuelle_1}
\end{equation}
where $\epsilon= 6R / (L v_0 \sqrt{bV_0})$, $V_0$ being the value of the voltage used for normalization. \\

Fig. S2 compares the space-time diagram recorded in the experiment (left column) with the space-time diagram converted to dimensionless units (right column) using the transformations specified by Eq. (\ref{normalization}). The initial condition is identical to the one used in the Letter: it is a soliton gas (SG) parameterized by $200$ discrete eigenvalues $\eta_k$ that are randomly distributed between $0.2$ and $0.6$ in an uniform way. After the step of the numerical spectral synthesis of the SG (see Sec. \ref{sec2}), it is generated by a programmable electric waveform generator having a bandwidth of $100$ MHz and a sampling rate of $1$ GS/s. The SG has an initial duration of  $\sim 28 \mu$s. The typical timescale associated with voltage fluctuations, which are linked to the randomness of the SG, is around $150$ ns. The experimental space-diagram is recorded by measuring the voltage every $5$ LC cells using an oscilloscope having a bandwidth of $200$ MHz.\\

The normalized time interval over which the evolution of the SG is observed is as large as $420$. Such an extended evolution time is achieved not in a single pass through the NLTL, but in $6$ passes. Each pass corresponds to a normalized time interval $\Delta t=70$, which is equivalent to physical propagation through the $160$ LC cells composing the NLTL. To achieve multiple passes in the NLTL, the electrical voltage recorded at the final LC cell after each pass is used as the initial condition for the subsequent pass.\\

Due to the small series resistance of the inductors ($\sim 0.09 \, \Omega$), there is slight damping in the NLTL. Consequently, the mass, defined as $M=\int u(x,t) dx$, is not perfectly conserved in a single pass. In the experiment, the mass decays by $\sim 5\%$ over $\Delta t=70$. This mass decay is compensated for after each pass by a slight amplification of the signal recorded at the final LC cell before it is reinjected in the NLTL. This compensation results in small periodic oscillations of the mass, as shown in Fig. S2(c). However, the mass is well conserved on average over the entire evolution of the SG between $t=0$ and $t=420$. \\

\begin{figure}[h]
    \includegraphics[scale=1]{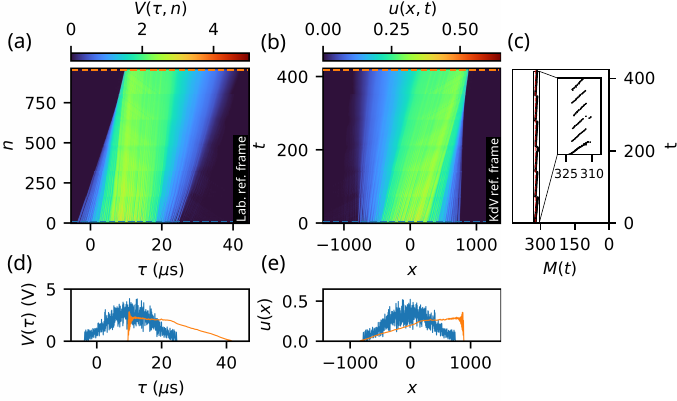}
    \caption{(a) Space-time diagram recorded in the NLTL by using a $200$ MHz oscilloscope. The voltage is measured every $5$ LC cells. $n$-index indicates the LC cell at which the voltage is probed. (b) Normalized experimental space-time diagram computed by using Eq. (\ref{normalization}). (c) Time evolution of the measured mass $M(t)=\int u(x,t) dx$ showing small oscillations due to the weak damping and its compensation at each pass in the NLTL. (d) Time evolution of the voltage measured at $n=0$ and $n=960$, after $6$ passes in the NLTL. (e) Normalized field $u(x,t)$ at $t=0$ and $t\sim 420$.}
    \label{normalization_procedure}
\end{figure}

\section{\label{sec2} Nonlinear spectral synthesis of soliton gases using the Darboux method}

The algorithm to generate N-soliton solution of the KdV equation with $N=200$ is based on the Darboux method. The Darboux transform is a recursive transformation scheme where the fundamental soliton solution of the KdV equation is used as a building block for the construction of higher-order $N$-soliton solutions through the iterative addition of new discrete eigenvalues $\eta_k$. The algorithm for the generation of KdV SG has been developed in ref. \cite{Congy:23}. Refering the reader to ref. \cite{Congy:23,Liao:96} for more details about the mathematical description of the Darboux machinary, we only provide here the algorithmic recipe to generate the N-soliton solution in a practical way.\\

- The first step consists is generating an ensemble of $N$ discrete eigenvalues $\eta_k$ ($k=1-N$) ranked in ascending order, from the smallest to the largest. An ensemble of $N$ position parameters $x_{0k}$ must also be created. For generating the SG used as initial condition in our experiment, the position parameters $x_{0k}$ are randomly and uniformely distributed over the interval $[-25,25]$.  The spectral parameters $\eta_k$ associated with this SG are randomly distributed between $0.2$ and $0.6$ in an uniform way. \\

- A numerical box of size $L$ and a given time $t$ must then be defined is such a way that the quantities:
\begin{equation}
  \Theta_k(x,t)=\eta_k(x-x_{0k}-4\eta_k^2 t)
\end{equation}
can be computed for $x \in [-L/2,L/2]$ for all the values of $k$ ($k=1-N$). \\

- For each $k$ between $1$ and $N$, compute : 
\begin{equation}
  \begin{aligned}
    & q_1(x,t,\eta_k) = \eta_k \tanh(\Theta_k(x,t)) \quad  \mathrm{ for \, \,  k \,  \,odd} \\
    & q_1(x,t,\eta_k) = \eta_k (\tanh(\Theta_k(x,t)))^{-1} \quad  \mathrm{for \,\, k \,\, even} \\    
  \end{aligned}    
\end{equation}

- For $N \geq k \geq 2$, compute:
\begin{equation}
  q_k(x,t,\eta_k)= \frac{\eta_k^2 -\eta_{k-1}^2}{q_{k-1}(x,t,\eta_k) - q_{k-1}(x,t,\eta_{k-1}) } - q_{k-1}(x,t,\eta_{k-1})
\end{equation}

- For $N \geq k \geq 2$, compute iteratively
\begin{equation}
  u_k(x,t)=  - u_{k-1}(x,t) - 2\eta_k^2 + 2 q_k(x,t,\eta_k)^2
\end{equation}
by starting the iteration from $u_1(x,t)=-2\eta_1^2 \, \text{sech}^2(\Theta_1(x,t))$. The N-soliton solution $u(x,t)$ of the KDV equation associated with the distribution of the spectral parameters $\eta_k$ initially chosen is obtained by changing $u_N(x,t)$ into $-u_N(x,t)$.

\section{\label{sec3} Solitary Wave Propagation, Mass Conservation, and Continuous Spectrum Excitation in the KdVB Equation}

\begin{center}
\fbox{%
  \begin{minipage}{0.9\textwidth}
The question of the evolution of a KdV soliton under the influence of dissipative perturbations has been extensively studied in the literature \cite{Madsen:69,Grimshaw:70,Johnson:73,Kaup:78,Kodama:81,Kivshar:89,Grimshaw:94,Johnson:94,Ko:78,Knickerbocker:80,Knickerbocker:85,GEl:02}. In particular, the concept that new discrete eigenvalues can be generated due to the excitation of continuous spectrum radiation has been discussed in references \cite{Kaup:78,Wright:80,Newell_book_solitons} in the context of solitary wave propagation in shallow water with slowly varying depth. In this section, we present numerical simulations to describe the long-term evolution of a solitary wave, aiming to synthesize and illustrate the nonadiabatic features previously identified in the literature within the framework of KdVB dynamics. 
\end{minipage}
}
\end{center}

\begin{figure}[t]
  \includegraphics[width=1.\textwidth]{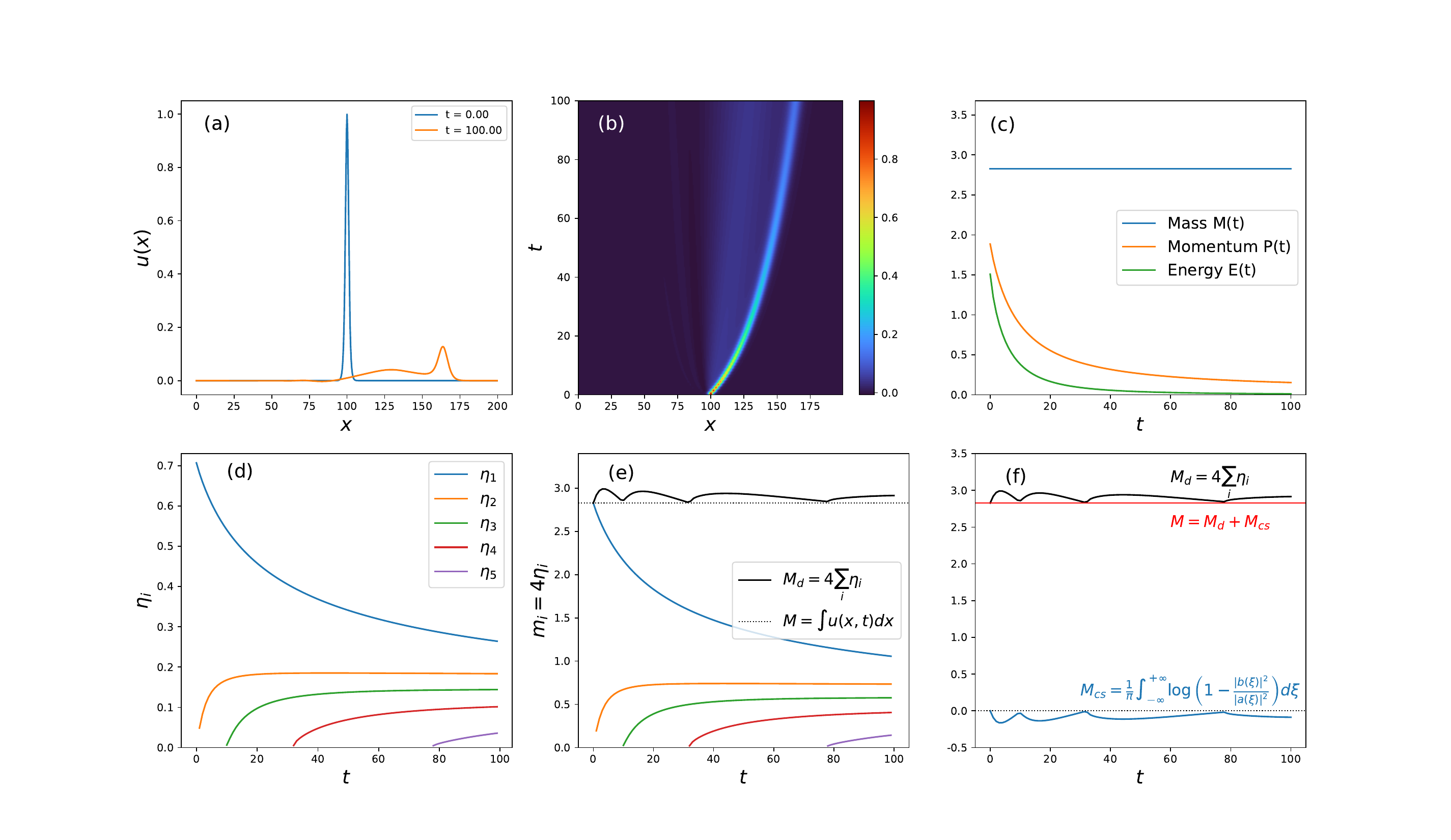}  
  \caption{Numerical simulations of Eq. (\ref{eq:kdvb_2}) with $\epsilon=0.1$. The initial condition is a soliton given by $u(x,t=0)=2\eta_1^2 \, sech^2(\eta_1 \,x)$ with $\eta_1=0.707$. (a) Field $u(x,t)$ numerically computed at $t=0$ (blue line) and at $t=100$ (orange line). (b) Space-time plot showing the evolution of the solitary wave. (c) Corresponding time evolution of the mass $M$, of the momentum $P$ and of the energy $E$. (d) Time evolution of the discrete eigenvalues successively created during the time evolution of the solitary wave. (e) Time evolution of the discrete masses defined as $m_i=4 \eta_i$ and of the discrete mass $M_{DS} = \sum \limits_{i=1}^5 4 \eta_i$. The dashed black line represents the total mass defined as $M=\int u(x,t) dx$. (f) Time evolution of the discrete mass $M_{DS}$ (black line) and of the mass of the continuous spectrum  defined as $M_{cs}=\frac{1}{\pi} \int_{-\infty}^{+\infty} \log \left( 1 - \frac{|b(\xi)|^2}{|a(\xi)|^2} \right) d\xi$. The red line is the total mass defined as $M=M_{DS}+M_{CS}$ according to Eq. (\ref{eq:mass_ist}). }
\end{figure}

\noindent
As in the Letter, we consider the KdVB equation
\begin{equation}\label{eq:kdvb_2}
  u_t + 6 u u_x +u_{xxx} = \epsilon u_{xx}, 
\end{equation}
and its associated Schr\"odinger equation in the IST theory
\begin{equation}\label{eq:schrodinger_2}
  \psi_{xx} + (u(x,t) -\lambda) \psi=0.
\end{equation}
The mass $M$ of the field $u(x,t)$ is defined as :
\begin{equation}\label{eq:mass_def}
  M=\int_{-\infty}^{+\infty} u(x,t) dx. 
\end{equation}  
The mass $M$ is a conserved quantity both in KdV and in KdVB equations: $dM/dt=0$. The momentum $P$ is defined as : 
\begin{equation}\label{eq:momentum_def}
  P=\int_{-\infty}^{+\infty} u^2(x,t) dx. 
\end{equation}  
The momentum $P$ is a conserved quantity in KdV equation ($dP/dt=0$) but it is not a conserved quantity in KdVB equation. For the KdVB equation, it is easy to show that :
\begin{equation}\label{eq:momentum_time}
  \frac{dP}{dt}= -2 \epsilon \int_{-\infty}^{+\infty} u_x^2(x,t) dx,  
\end{equation}  
which means that the rate at which this quantity decays is determined by the gradients $u_x$ of the field in space.
The energy $E$ is defined as : 
\begin{equation}\label{eq:energy_def}
  E=\int_{-\infty}^{+\infty} \left( u^3(x,t) - \frac{u_x^2(x,t)}{2} \right) dx. 
\end{equation}
The energy is a conserved quantity in KdV but not in KdVB. In KdVB, the energy changes in time but does not necessarily decrease.

In the IST formalism, the mass $M$ and the momentum $P$ have the following expressions:
\begin{equation}\label{eq:mass_ist}
  M= 4 \sum_k \eta_k + \frac{1}{\pi} \int_{-\infty}^{+\infty} \log \left( 1 - \frac{|b(\xi)|^2}{|a(\xi)|^2} \right) d\xi
\end{equation}  
\begin{equation}\label{eq:momentum_ist}
  P= \frac{16}{3} \sum_k \eta_k^3 - \frac{4}{\pi} \int_{-\infty}^{+\infty} \xi^2 \log \left( 1 - \frac{|b(\xi)|^2}{|a(\xi)|^2} \right) d\xi
\end{equation}  
where $\eta_k$ are the discrete eigenvalues that can be computed by solving Eq. (\ref{eq:schrodinger_2}) for the potential $u(x,t)$. $\rho(\xi)=b(\xi)/a(\xi)$ is the reflection coefficient defined in the IST theory, see e.g. ref. \cite{Kaup:78,Wright:80,Newell_book_solitons}. In Eq. (\ref{eq:mass_ist}) (resp. Eq. (\ref{eq:momentum_ist})), the discrete summation over the $\eta_k$ (resp. $\eta_k^3$) is related to the discrete part of the spectrum. On the other hand the integral term in Eq. (\ref{eq:mass_ist}) (resp. (\ref{eq:momentum_ist})) represents the contribution to the mass (resp. to the momentum) of the continuous part of the spectrum that is associated with the radiative modes. \\

Fig. S3 shows numerical simulations of Eq. (\ref{eq:kdvb_2}) with $\epsilon=0.1$ for an initial condition being under the form of a soliton given by $u(x,t=0)=2\eta_1^2 \, sech^2(\eta_1 \,x)$ with $\eta_1=0.707$, like in the experiment shown in Fig. 2 of the Letter. The maximum evolution time reached in the numerical simulation shown in Fig. S3 is $100$, which is significantly larger than the experimental evolution time ($t \simeq 16$). The presence of diffusion described by the term on the right-hand side of Eq. (\ref{eq:kdvb_2}) induces a slow decay in the soliton's amplitude and a gradual decrease in its velocity, as illustrated in Fig. S3(a)(b). This evolution is accompanied by the formation of a shelf and of small oscillations behind the solitary wave. The mass $M$ is conserved during the evolution while the momentum $P$ and the energy $E$ decrease over time, see Fig. S3(c).   \\

Fig. S3(d) shows that the amplitude $\eta_1$ of the discrete eigenvalue linked to the initial condition gradually decays over time, while $4$ new discrete eigenvalues are created during the shelf formation. Fig. S3(d) shows the evolution of the masses $m_i=4\eta_i$ of each of the $4$ discrete eigenmodes, see Eq. (\ref{eq:mass_ist}). Of course the discrete masses $m_i(t)=4\eta_i(t)$  $(i=1-4)$ follow the same time evolution as the discrete eigenvalues $\eta_i(t)$. The interesting point is that the sum of the discrete masses $M_{DS}(t) = \sum \limits_{i=1}^8 4 \eta_i(t)$ is not constant but slightly fluctuates above the constant value of the total mass $M=\int u(x,t) dx$, see the full and dotted black lines in Fig. S3(e). \\

The decay in time of $\eta_1$ is associated to a decrease in mass, which must be compensated by an increase to preserve the fact that the total mass $M=\int u(x,t) dx$ must be conserved \cite{Kivshar:89}. The missing mass is provided by the generation of the new discrete eigenvalue $\eta_2$. However the decay of $\eta_1(t)$ and the growing of $\eta_2(t)$ can never be balanced in such an exact way that the mass associated with these two discrete eigenvalues is constant in time : $m_{1,2}(t)=m_1(t)+m_2(t)=4(\eta_1(t)+\eta_2(t)) \neq cte$. For the total mass conservation condition $dM/dt=0$ to be fulfilled, there is no choice but for the mass $M_{cs}$ associated with the radiative modes to fluctuate in the same way as $M_d$ but with an opposite sign ($M_{CS}=\frac{1}{\pi} \int_{-\infty}^{+\infty} \log \left( 1 - \frac{|b(\xi)|^2}{|a(\xi)|^2} \right) d\xi$). This is illustrated in Figure S3(f), which shows that the negative mass fluctuations of the continuous spectrum exactly compensate for the positive fluctuations of the discrete mass $M_{DS}$, so that the total mass $M = M_{DS} + M_{CS}$ is constant over time.\\


\section{\label{sec4} Soliton condensates of genus 0 and genus 1 }

The notion of soliton condensate as the ‘densest possible soliton gas’ for a given spectral support was introduced in \cite{GEl:20} in the context of the focusing NLS equation. A detailed theory of soliton condensates for the KdV equation has been developed in \cite{Congy:23}. Here, we refer the reader to these papers and only provide the mathematical expressions for the densities of state (DOS) of soliton condensates of genus $0$ and $1$. We also show that these mathematical expressions fit very well the DOS found in numerical simulations reported in Fig. 3 of the Letter. \\

\begin{figure}[h]
  \includegraphics[width=0.5\textwidth]{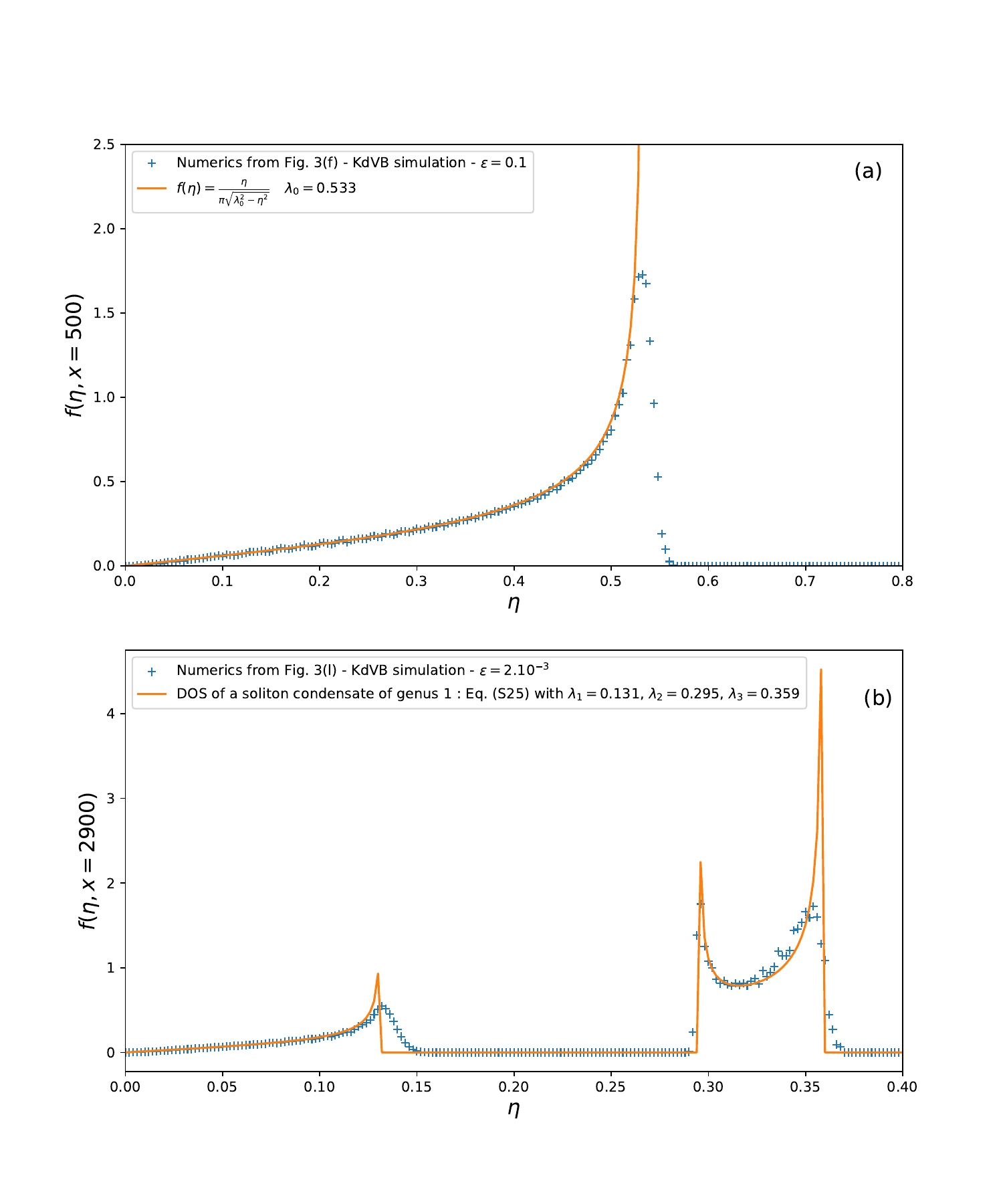}
  \caption{
    (a) The blue crosses represents the spectral distribution shown in the inset of Fig. 3(f). The orange line fits the blue crosses with the model (\ref{eq:weyl}) with $\lambda_1=0.533$. (b) The blue crosses represents the spectral distribution shown in the inset of Fig. 3(l). The orange line fits the blue crosses with the model  (\ref{eq:genus1}) with $\lambda_1=0.131$, $\lambda_2=0.295$, $\lambda_3=0.359$.
   }
\end{figure}

The DOS of a KdV soliton condensate of genus $0$ is given by the so-called Weyl distribution that reads: 
\begin{equation}\label{eq:weyl}
  f(\eta)=\frac{\eta}{\pi \sqrt{\lambda_1^2-\eta^2}}; \qquad \eta \in [0,\lambda_1].
\end{equation}  
Fig. S4(a) shows that the DOS of the SG measured at $t=420$ in the electrical NLTL ($\epsilon=0.1$) is very well fitted by the Weyl distribution with $\lambda_1=0.533$, see also Fig. 3(f) in the Letter.\\

The DOS of a KdV soliton condensate of genus $1$ lives on a spectral support not composed of one band but of two bands with their endpoints being given by the three real parameters $\lambda_1$, $\lambda_2$, $\lambda_3$. The DOS of the soliton condensate of genus 1 reads
\begin{equation}\label{eq:genus1}
  f(\eta)= \frac{ i \eta (\eta^2 - w^2)}{ \pi \sqrt{ (\eta^2 - \lambda_1^2) (\eta^2 - \lambda_2^2) (\eta^2 - \lambda_3^2) }};   \qquad \eta \in [0,\lambda_1] \cup [\lambda_2,\lambda_3]
\end{equation}
with $w^2=\lambda_3^2-(\lambda_3^2-\lambda_1^2) \chi(m)$, $\chi(m)=E(m)/K(m)$ and $m=(\lambda_2^2-\lambda_1^2)/(\lambda_3^2-\lambda_1^2)$. $E(m)$ (resp. $K(m)$) is the complete elliptic integral of the second (resp. first) kind.

Fig. S4(b) shows that the DOS of the SG measured at ($x=2900$, $t=4000$) in the numerical simulations of the KdvB equation with small diffusion ($\epsilon=2.10^{-3}$) is remarkably well fitted by Eq. (\ref{eq:genus1}) with $\lambda_1=0.131$, $\lambda_2=0.295$, $\lambda_3=0.359$, see also Fig. 3(l) in the Letter. It was conjectured and numerically verified in \cite{Congy:23} that realizations of the genus $1$ condensate with DOS (\ref{eq:genus1}) almost surely coincide with periodic wave solutions of KdV parameterized by $\lambda_1$, $\lambda_2$ and $\lambda_3$. These solutions locally describe diffusive-dispersive shock waves of the KdVB equation \cite{GEl:17}.

\end{document}